\documentclass[conference]{IEEEtran}
\IEEEoverridecommandlockouts

\AtBeginDocument{%
	}

\usepackage{comment}
\usepackage{url}
\usepackage{subcaption}
\usepackage{pifont}
\usepackage{tikz}
\usepackage{fontawesome5}
\usepackage{arabtex}
\usepackage{utf8}
\usepackage{xurl}
\usepackage{fdsymbol}
\usepackage{booktabs}
\usepackage{flushend}
\setcode{utf8}

\usepackage[skipabove=5pt,skipbelow=5pt,roundcorner=10pt,framemethod=TikZ]{mdframed}

\definecolor{vlightgray}{gray}{0.925}
\newcommand{\descr}[1]{\smallskip\noindent\textbf{#1}}
\newcommand{\pval}{{0.01}}

\begin{document}

\author{Mohammad Hammas Saeed$^{\clubsuit}$, Shiza Ali$^{\clubsuit}$, Pujan Pauel$^{\clubsuit}$, Jeremy Blackburn$^{\vardiamondsuit}$, and Gianluca Stringhini$^{\clubsuit}$\\[0.5ex]
 $^{\clubsuit}$Boston University, $^{\vardiamondsuit}$Binghamton University\\
\normalsize \{hammas,shiza,ppaudel,gian\}@bu.edu, jblackbu@binghamton.edu\vspace*{-0.3cm}}\date{}

\title{Unraveling the Web of Disinformation: Exploring the Larger Context of State-Sponsored Influence Campaigns on Twitter\thanks{This paper is accepted for publication in the Proceedings of the 2024 International Symposium on Research in Attacks, Intrusions and Defenses (RAID). Please cite accordingly.}}

\maketitle

\begin{abstract}
	Social media platforms offer unprecedented opportunities for connectivity and exchange of ideas; however, they also serve as fertile grounds for the dissemination of disinformation.
	Over the years, there has been a rise in state-sponsored campaigns aiming to spread disinformation and sway public opinion on sensitive topics through designated accounts, known as \textit{troll accounts}.
	Past works on detecting accounts belonging to state-backed operations focus on a single campaign.
	While campaign-specific detection techniques are easier to build, there is no work done on developing systems that are campaign-agnostic and offer generalized detection of troll accounts unaffected by the biases of the specific campaign they belong to.
	
	In this paper, we identify several strategies adopted across different state actors and present a system that leverages them to detect accounts from previously unseen campaigns.
	We study 19 state-sponsored disinformation campaigns that took place on Twitter, originating from various countries.
	The strategies include sending automated messages through popular scheduling services, retweeting and sharing selective content and using fake versions of verified applications for pushing content.
	By translating these traits into a feature set, we build a machine-learning-based classifier that can correctly identify up to 94\% of accounts from unseen campaigns.
	Additionally, we run our system in the wild and find more accounts that could potentially belong to state-backed operations.
	We also present case studies to highlight the similarity between the accounts found by our system and those identified by Twitter.
\end{abstract}

\section{Introduction}
Social media platforms have become prominent sources for accessing information and communication for millions of people worldwide.
As these platforms are used more and more for information propagation, there has been an associated risk of \textit{misinformation} and \textit{disinformation}.
\textit{Misinformation} is the spread of false information without malicious intent (e.g., a regular Twitter user innocuously promotes a COVID-19 false narrative tweet), whereas \textit{disinformation} is the intentional spread of false information by malicious actors~\cite{shao2018anat, lisa2020covid, starbird2017examining, vosoughi2018spread}.
Recent years have seen a rise in state-sponsored disinformation campaigns, where governments and their affiliated entities exploit social media platforms to shape narratives, manipulate public opinion, and advance their strategic agendas~\cite{starbird2017examining, starbird2019disinformation}.
These campaigns are often conducted by a designated set of accounts, known as \emph{troll accounts}, which are created by malicious actors and often manually controlled to post content and interact with each other and real users~\cite{marco2019brexit, bessi2016social, ferrara2016rise, shao2018bots, varol2017online}.
The actors, often operating covertly, exploit various channels and tactics to disseminate misleading information.
For example, in 2020, a six-year-long Russian disinformation campaign named ``Secondary Infektion'' was found to be spreading pro-Russian narratives and interfering with the 2016 US presidential election across 300 social media platforms in seven different languages~\cite{study2020bobby}.

\descr{Motivation.} Over the years, disinformation campaigns have grown in scale and are now operating globally.
According to recent studies, over 200 operations targeting various countries were taken down by Meta in 2022 alone~\cite{martin2022meta}.
In fact, many campaigns are now being outsourced to third-party agencies (e.g., troll-farms~\cite{sheb2019craft} and PR firms~\cite{stan2020pr}) by political actors~\cite{gold2021disinf}.
However, unlike social media bots and spam, these campaigns are often more sophisticated and deliberate in nature~\cite{saeed2022troll}, making the threat model more intricate.
The bare-bones attack model is often two-fold: 1) the campaign is geared towards achieving a certain goal (e.g., spreading an ideology, causing conflict and strife) on a target platform, while 2) slipping under the radar of detection systems, blending into the community, and appearing ``legitimate.''
Since these campaigns exhibit group activity and human-like behavior~\cite{ezzexpose2023, saeed2022troll}, it is important to develop systems that are specifically tailored towards them and more sophisticated than systems designed to detect automated activity.

With the passage of time, various evasion tactics are being used by troll accounts to protect from detection systems and human moderation~\cite{lew2017misinf, shao2018spread, zannettou2019web}.
Despite this, little work has been done to understand the shared characteristics of these campaigns to develop systems that can offer broad-scale generalized detection.
Some recent research has leaned towards understanding the common traits of these campaigns; e.g., miscreants purposely pump a large quantity of ``spammy'' comments on a target platform to divert attention from the original narratives being pushed, or designated fake accounts within a campaign are created for specific tasks~\cite{martin2022meta}.
Other works have looked into specific state-sponsored campaigns or events on Twitter (e.g., the 2018 Brazilian presidential campaign~\cite{felipe2021disinf}, Italian disinformation during the 2019 European elections~\cite{pierri2020disinf}, and Internet Research Agency (IRA) accounts~\cite{yiping2019disinf}).

\descr{Technical Roadmap.} Unlike past research that is campaign-specific, our research focuses on uncovering the common themes among disinformation campaigns to perform campaign-agnostic detection. 
In this paper, we identify several universal traits that are shared among different campaigns.
We analyze Twitter data for state-sponsored campaigns that spans 19 countries and includes more than 200 million tweets from the years 2018 to 2022.
We find that these campaigns use a variety of techniques to perform their operations, e.g., using scheduling services to delegate their posting tasks, utilizing fake third-party versions of popular applications (e.g., ``Twitter for Android'') to post messages, extensively retweeting to push certain agendas, and posting innocuous messages (e.g., motivational quotes) to potentially avoid detection.
We also highlight potential coordination patterns where accounts from different campaigns exhibit similar characteristics (e.g., using the same Twitter sources to post their messages) around the same timeframe, making our findings in line with some recent works pointing towards potential inter- and intra-state coordination in campaigns~\cite{wang2023state}.

Overall, we identify several universal traits and create a cross-campaign detection system that can detect upto 94\% accounts from unseen campaigns.
We demonstrate the efficacy of our system by training the classifier on each campaign one by one and detecting accounts from all the other campaigns.
Lastly, we find potential malicious accounts in the wild and highlight some case studies that showcase their involvement in disinformation campaigns.

\descr{Contributions.} This paper makes the following contributions:

\begin{itemize}
\item We uncover salient features used by state-sponsored disinformation operations (e.g., using fake third-party applications). These findings shed light on ways to identify this activity and detect potential malicious accounts.

\item We develop a machine learning classifier to detect malicious accounts from previously unseen campaigns. Our most successful implementation uses a Random Forest model with an F1-Score of 97.8\%. We also perform cross-campaign detection, flagging upto 94\% accounts from unseen campaigns.

\item We evaluate our system on 2,696 Twitter accounts using fake third-party applications. We identify 116 new malicious accounts potentially belonging to state-backed operations and present case studies to provide evidence that these accounts operate similarly to accounts from known campaigns.
\end{itemize}

\descr{Paper Organization.}
The rest of the paper is organized as follows.
The next section describes our dataset.
In Section 3, we analyze the campaigns in our dataset and highlight the common strategies they use, which informs the development of our detection model.
Next, we translate our findings into features to build a machine learning classifier that can distinguish between real accounts and state-backed accounts in Section 4, while in Section 5, we show that our system can perform cross-campaign detection.
In Section 6, we discuss several case studies of accounts that our system detects in the wild, while in Section 7, we discuss other works related to our research; finally, we discuss the implications of our results, future work, and conclude the paper in Section 8.

\begin{table}[t]
	\begin{center}
		\setlength{\tabcolsep}{3.5pt}
		\small
		\begin{tabular}{ll}
			\toprule
			\textbf{Campaign}  &  \textbf{Number of Accounts}   \\ 
			\midrule
			2019nov\_saudia & 
			5,929 \\
			2019aug\_china2 &
			4,301 \\
			2019aug\_uae &
			4,248 \\
			2018oct\_ira &
			3,613 \\
			2019jun\_iran2 &
			2,865 \\
			2019jan\_iran &
			2,320 \\
			2019jan\_venezuela1 &
			1,196 \\
			2019jun\_iran &
			1,666 \\
			2019aug\_ecuador &
			1,019 \\
			2018oct\_iran &
			770 \\
			2019jan\_venezuela &
			764 \\
			2019aug\_china &
			744 \\
			2019jan\_russia &
			416 \\
			2019jun\_iran1 &
			248 \\
			2019aug\_china1 &
			197 \\
			2019jun\_catalonia &
			130 \\
			2019jun\_venezuela &
			33 \\
			2019jan\_bangladesh&
			15 \\
			2019jun\_russia & 
			4 \\
			\bottomrule
		\end{tabular}%
	\end{center}
	\caption{Number of Accounts Per Campaign.}
	\label{tbl:data}
\end{table}

\begin{figure*}
	\centering
	
	\begin{subfigure}[b]{0.49\textwidth}
		\centering
		\includegraphics[width=\textwidth]{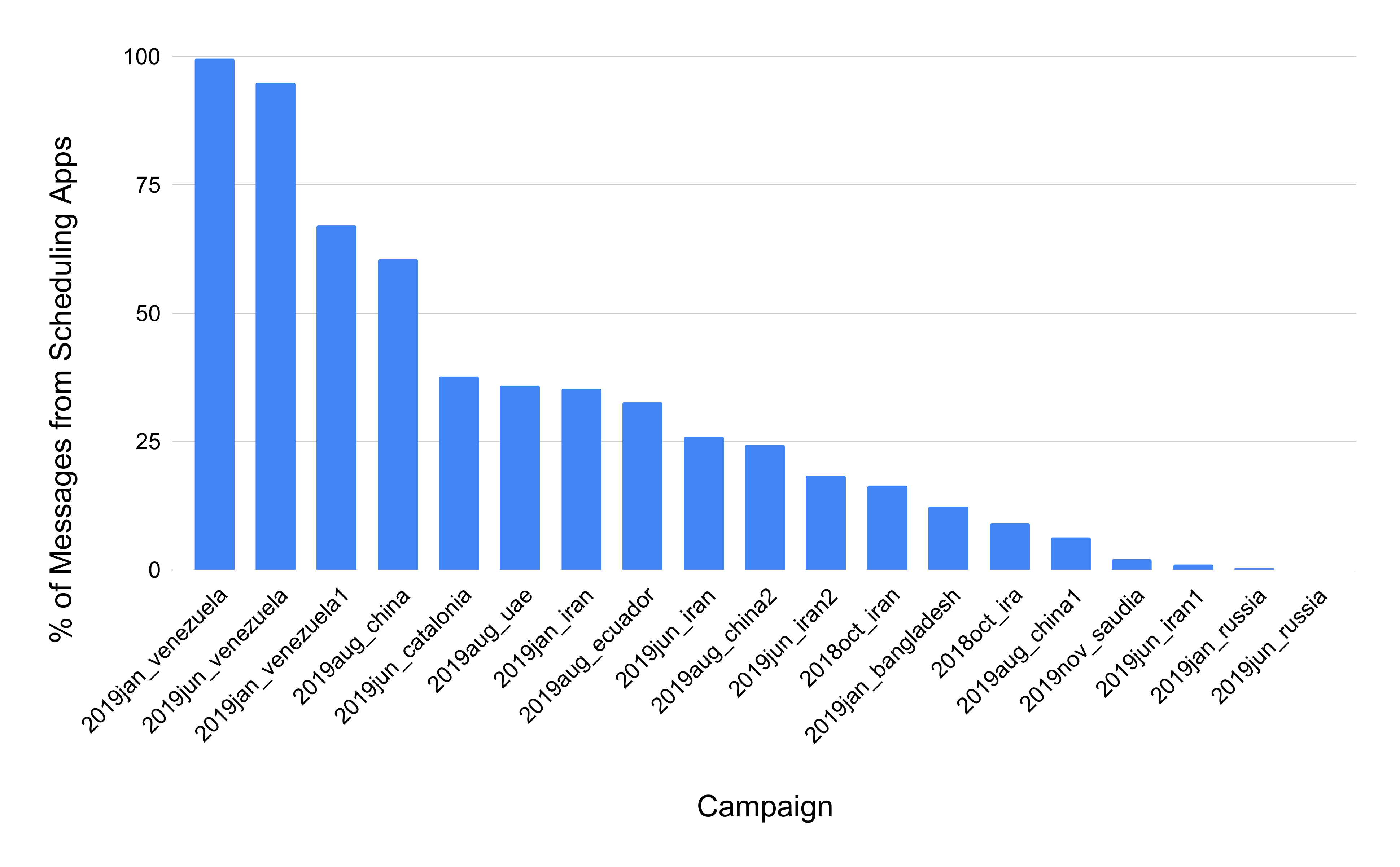}
		\caption{Scheduled Messages}
		\label{fig:sched}
	\end{subfigure}
	\hfill
	\begin{subfigure}[b]{0.49\textwidth}
		\centering
		\includegraphics[width=\textwidth]{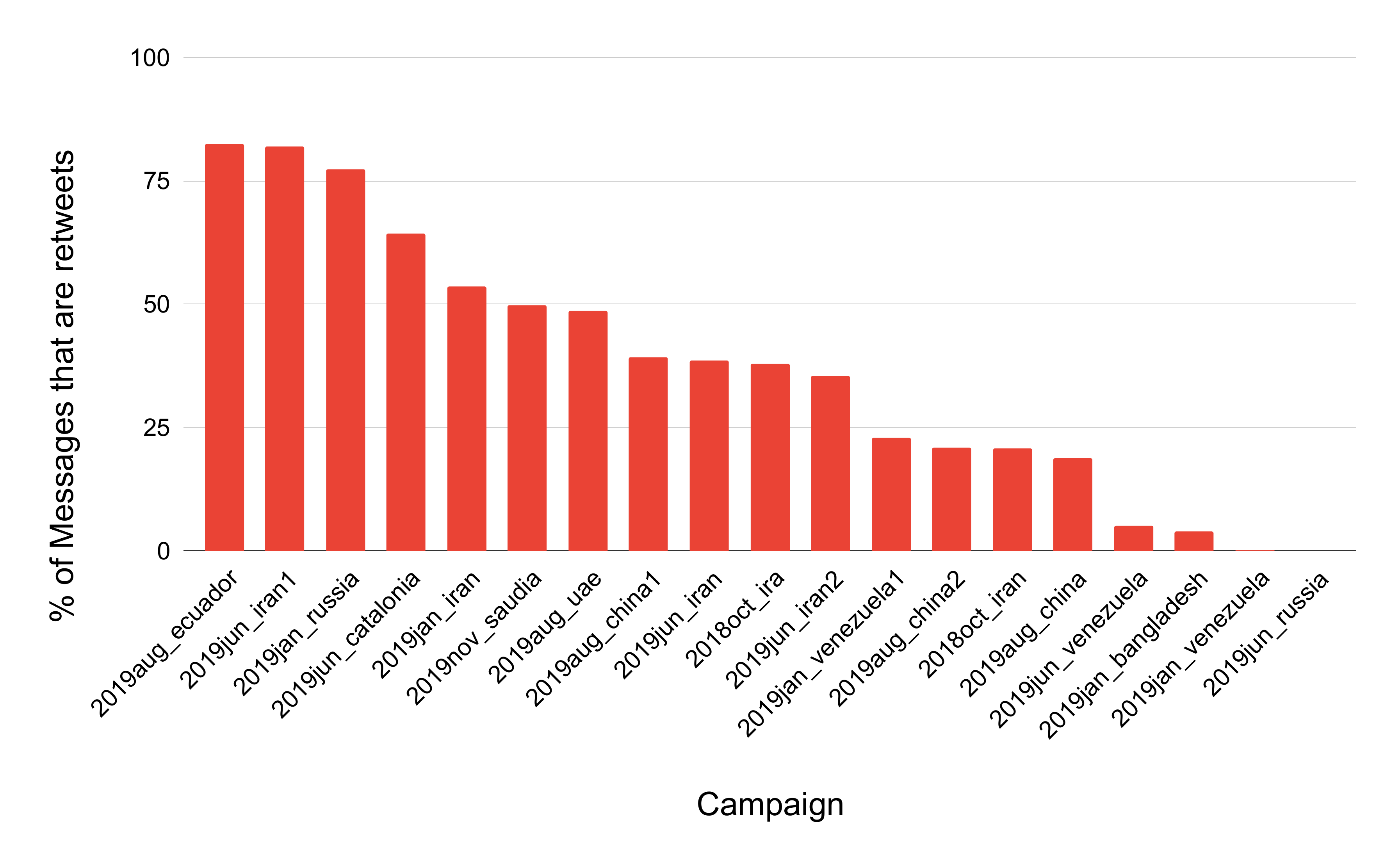}
		\caption{Retweeted Messages}
		\label{fig:retweet}
	\end{subfigure}
	
	\caption{The graphs show the percentage of messages through (a) scheduling applications and (b) the percentage of messages that are retweets.}
	\label{fig:techniques}
\end{figure*}

\section{Data}\label{sec:data}
We use Twitter for our analysis, as the platform released a publicly available dataset from campaigns active in diverse countries.
Twitter is a popular social media platform that enables users to share and interact with short messages, known as tweets, in real-time.
With its widespread adoption and influence, Twitter has attracted attention from researchers across various domains, with past works exploring different aspects of Twitter, including user behavior~\cite{cheng2014can}, information diffusion~\cite{bakshy2011everyone}, and fake account detection~\cite{galangarcia, stringhini2010detecting}.
Later on, we use Twitter's API for collecting tweets; note that the Twitter free API was discontinued during the course of this project, and we discuss its implications in Section~\ref{sec:lim}.
Broadly, we divide our dataset into two categories:

\descr{Campaign Data.}
The Twitter Transparency Dataset is a collection of publicly available information provided by Twitter related to various aspects of the platform's operations.
It aims to promote transparency and enable researchers, journalists, and the public to analyze and understand Twitter's activities, policies, and content moderation efforts.
We use the data from the Twitter Moderation Research Consortium~\cite{twittercamps} which contains information on platform manipulation campaigns attributed to various state-backed actors from 2018 to 2022.
We study 19 campaigns, spanning more than 200 million Tweets and nine terabytes of media.
It contains data from almost 80,000 accounts in total.
Table~\ref{tbl:data} shows the campaigns we use in our dataset and the number of accounts in each campaign.
Throughout the paper, the naming convention we use for campaigns is: \{year\}\{month\}\_\{country\}\{variant\}.
The month and year refer to the time when said campaign was \emph{detected} and is officially designated by Twitter in the dataset.
The variant portion of the name helps distinguish between campaigns because there are multiple active campaigns from the same country detected at similar time periods.

\descr{Twitter 1\% Data.}
For this research, we use pre-crawled data from the Twitter 1\% Data API for the years 2017-2022.
The API used a 1\% random sample of public Tweets for a given day.
We use different parts of this data in later sections for finding regular accounts to compare with state-backed accounts.

\descr{Ethics.}
Our work is not categorized as human subjects research by our IRB since we do not interact with human subjects and solely use data that is already available to the public.
Nonetheless, we adhere to ethical standards by removing any personally identifiable information when presenting example tweets in the paper and redacting usernames from tweets to prevent deanonymizing users.

\section{Characteristics of Campaigns}\label{sec:charact}
In this section, we dive deeper into the different techniques used by state-sponsored campaigns to varying degrees.
This section is divided into two parts: 1) Campaign-Level characteristics and 2) Account-Level characteristics.
We first identify campaign-level characteristics by analyzing the tweets across various metrics like timing, coordination, content, and posting habits.
We also uncover patterns that are shared across campaigns.
We then map these campaign-level characteristics to account-level features, which can later aid us in developing a system to distinguish between real-world users and accounts from state-sponsored campaigns.

\subsection{Campaign-Level Characteristics}

\descr{Scheduled Messages. }
On Twitter, each tweet has a ``source'' field that refers to the application or platform from which the tweet was posted.
We analyze the source field of all tweets posted by disinformation campaigns and pick the top 50 most commonly used applications.
We identify seven popular scheduling applications (i.e., IFTTT, TweetDeck, dlvr.it, Hootsuite, Twibble, SocialOomph, and Zapier.com) among the top applications.
These applications are used to send a total of 15,600,226 messages across 19 campaigns.
In Figure~\ref{fig:sched}, we show that most coordinated campaigns use scheduling applications for a large percentage of their messages, with some campaigns posting almost all messages from scheduling applications, like the June 2019 and January 2019 Venezuelan campaigns.
On average, 30\% of the messages in a campaign belong to one of the scheduling applications.

\descr{Retweeting Similar Messages.}\label{sec:retweet}
A retweet allows users to share someone else's tweet with their own followers.
When a user retweets a tweet, it appears on their own timeline and is visible to their followers.
This enables users to amplify and redistribute content that they find interesting, informative, or worthy of sharing for other reasons.
We find that campaigns have upto 78\% of their entire tweets as retweets, as shown in Figure~\ref{fig:retweet}.
We also find that certain messages are retweeted both intra- and inter-campaign, which can be suggestive of a collaborative effort or a shared modus operandi (e.g., retweeting popular tweets to appear ``legitimate'').
For example, the following tweets are retweeted by multiple campaigns (i.e., 2019aug\_china1, 2019nov\_saudia, 2019aug\_ecuador and 2019aug\_china2) which are unrelated suggesting group-like behavior:

\begin{mdframed}[style=exampledefault, backgroundcolor=vlightgray]
	RT [REDACTED]: follow everyone that retweets this. \ding{34}
	
	\noindent RT [REDACTED]: Now that’s a t-shirt cannon if I've ever seen one 
	
	\noindent RT [REDACTED]: follow everyone who LIKES and RTs this \faIcon{fish}
	
	\noindent RT [REDACTED]: Need to mass unfollow? Go to \url{http://www.iunfollow.com} There are no limits and it's free! No signup required!
	
	\noindent RT [REDACTED]: \url{https://t.co/dTdDjRvXbs}
	
	\noindent RT [REDACTED]: Siga todos que FAV e RT esse tweet \faCircle
	
\end{mdframed}

These retweets are encouraging users to follow or like a particular account or tweet.
Similar patterns are observed in follower markets; services that use accounts to build fake influence and reputation.
Previous research extensively studied follower markets~\cite{stringhini2012poultry, stringhini2013follow,weerasinghe2020pod} and the detection of accounts involved in such activity~\cite{wu2014detect}.
However, it is important to note that follower-market tweets are not the only kind of retweets these accounts make; we exclude other examples for brevity.

\begin{figure}[!t]
	\centering
	\includegraphics[width=0.4\textwidth]{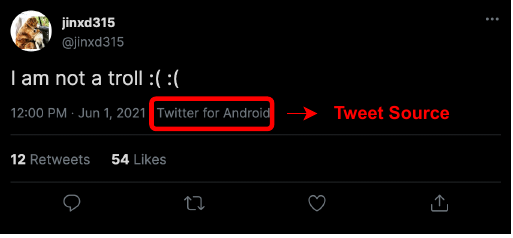}
	\caption{Tweet from ``Twitter for Android'' source.}
	\label{fig:tweetexample}
\end{figure}

\descr{Impersonating Third-Party Sources.}
We find that coordinated campaigns rely on specific third-party sources for posting content.
Many of these sources are fake versions of existing applications, since Twitter forces each application name to be unique.
For example, the fake version of ``Twitter{\textvisiblespace}for{\textvisiblespace}Android'' will be ``Twitter{\textvisiblespace}for\textvisiblespace{\textvisiblespace}Android'' with an extra space ({\textvisiblespace}) between ``for'' and ``Android.''
In Table~\ref{tbl:fakeapps}, we provide several examples of original sources alongside their corresponding fake sources, along with their frequency of occurrence. 
Upon manual inspection and from the examples in Table~\ref{tbl:fakeapps}, we conclude that fake sources often contain spelling errors, leading spaces, and other similar changes from the original app source name that are difficult to distinguish visually.
We search the latest Twitter 1\% Data from 2022, which contains a random 1\% sample of tweets made on each day of 2022, to find out if regular users also post with these sources.
Of all the applications listed in Table~\ref{tbl:fakeapps}, we only find two messages from ``\textvisiblespace Twitter for iOS,'' indicating these sources are not used by regular users and are more likely to be used by state-backed accounts.
Table~\ref{tbl:fakeapps} shows that an impersonated version of ``Twitter for Android'' was used to send 106,636 tweets.
State-sponsored accounts might use fake third-party sources for a variety of reasons.
For example, these sources might aid in the management of accounts by offering automation, added functionalities, and extra features that prove beneficial for the campaigns.
Another reason to use such sources is their legitimate-looking names, which make the account look legitimate at a cursory glance by users on the platform.
Figure~\ref{fig:tweetexample} shows how the source used to appear on Twitter and an extra space or basic string manipulation can go easily unnoticed.
We also did not observe regular users using these applications in our dataset.
Additionally, we search Google Play Store and Apple App Store and do not find these impersonated third party applications, indicating that these are not publicly available apps and are used exclusively by malicious actors.

\begin{table}[t]
	\begin{center}
		\setlength{\tabcolsep}{3.5pt}
		\small
		\begin{tabular}{lll}
			\toprule
			\textbf{Original Application}  &  \textbf{Third-Party Version(s)} & \textbf{Frequency} \\ \midrule
			``Twitter for Android'' & ``Twitter for\textvisiblespace\textvisiblespace Android'', & 106,636 \\ &
			 ``Twidere for Android \#2'', & 34 \\ &
			 ``Twidere for Android \#5'', & 54 \\ &			
			 ``Twidere for Android \#7'', & 73 \\ &
			``Twitter from Android'' & 3,283 \\&
			``android apps for twitter'' & 1,555\\ 
			``Twitter for iPad'' & ``Twitter for\textvisiblespace\textvisiblespace iPad'', & 20 \\ &
			 ``twtkr for iPad'', & 275 \\ &
			 ``\textvisiblespace\textvisiblespace\textvisiblespace\textvisiblespace twtkr for iPad'' & 73 \\
			``Twitter for iPhone'' & ``Twitter for iphons'', & 12,523\\ &
			``twitter for Iphone ios'' & 782\\ 
			``Instagram''      & ``\textvisiblespace\textvisiblespace\textvisiblespace Instagram''  & 3 \\   
			``Twitter for iOS'' & `` Twitter for iOS'' & 364\\  
			``HTC Peep''      & ``\textvisiblespace\textvisiblespace HTC Peep'' & 427 \\
			``Hootsuite'',    & ``hootsuite'', & 6,127 \\
			``Hootsuite Inc." & ``\textvisiblespace\textvisiblespace\textvisiblespace\textvisiblespace\textvisiblespace\textvisiblespace\textvisiblespace\textvisiblespace HootSuite'' & 43 \\     			       						
			\bottomrule
		\end{tabular}%
	\end{center}
	\caption{Third-party versions of original applications used by state-sponsored campaigns.}
	\label{tbl:fakeapps}
\end{table}

\begin{figure}[!t]
	\centering
	\includegraphics[width=0.5\textwidth]{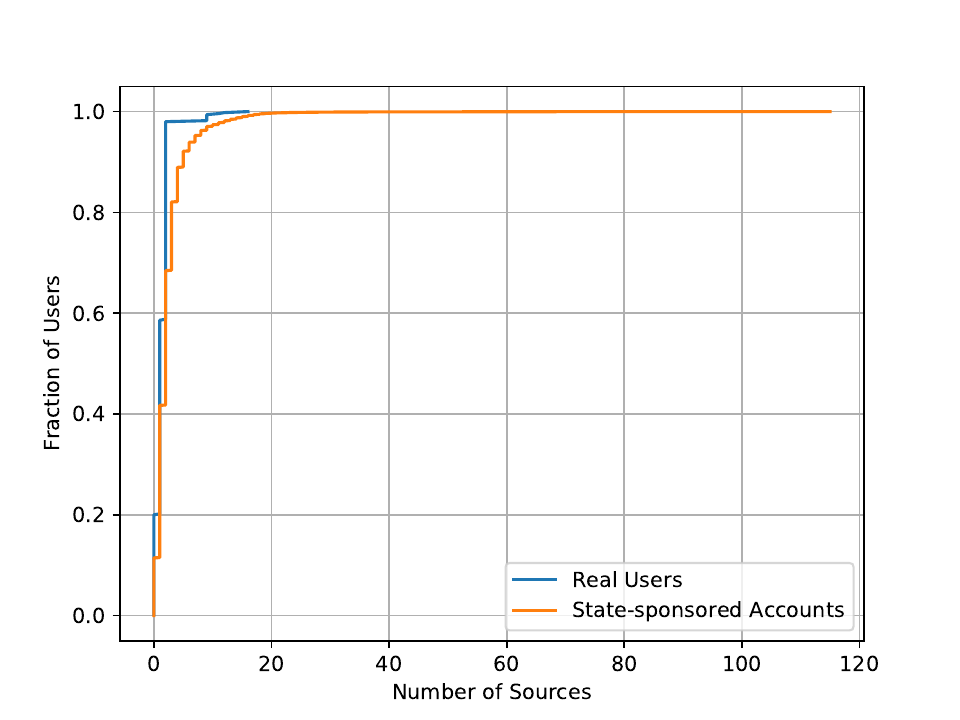}
	\caption{CDF of number of sources used by real users and accounts from all campaigns.}
	\label{fig:sources_cdf}
\end{figure}

\begin{figure*}
	\centering
	
	\begin{subfigure}[b]{0.49\textwidth}
		\centering
		\includegraphics[width=\textwidth]{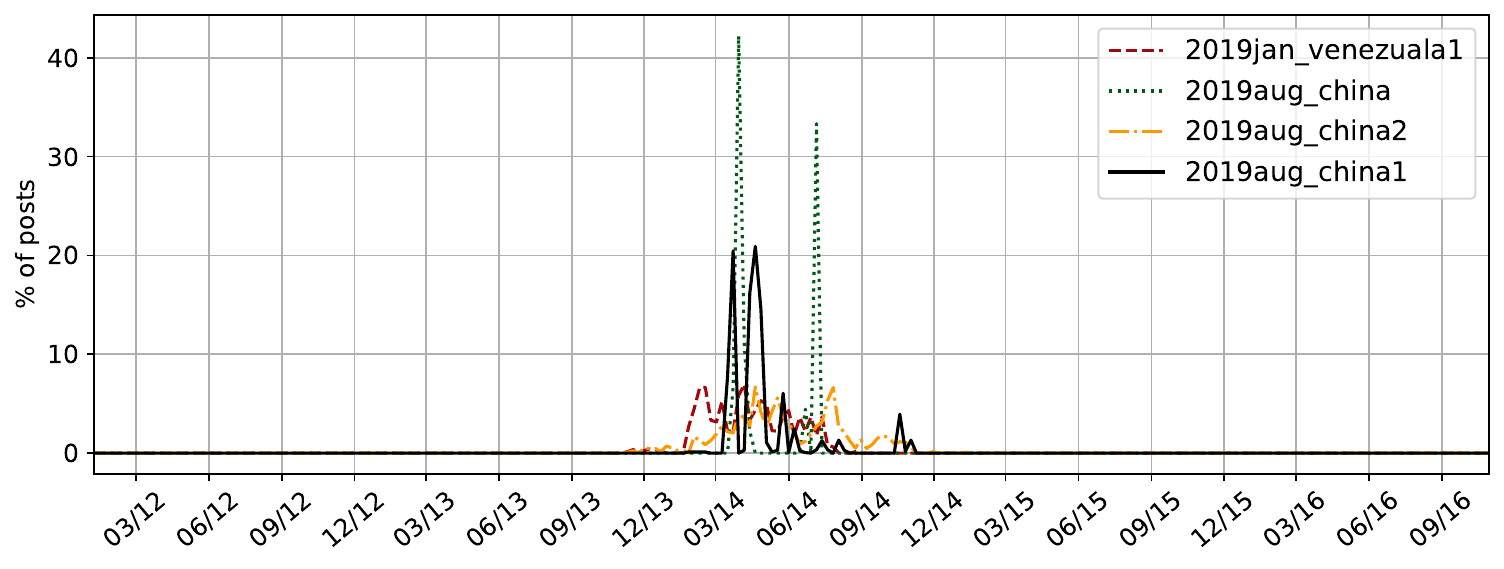}
		\caption{``Twitter for\textvisiblespace\textvisiblespace Android''}
		\label{fig:android}
	\end{subfigure}
	\hfill
	\begin{subfigure}[b]{0.49\textwidth}
		\centering
		\includegraphics[width=\textwidth]{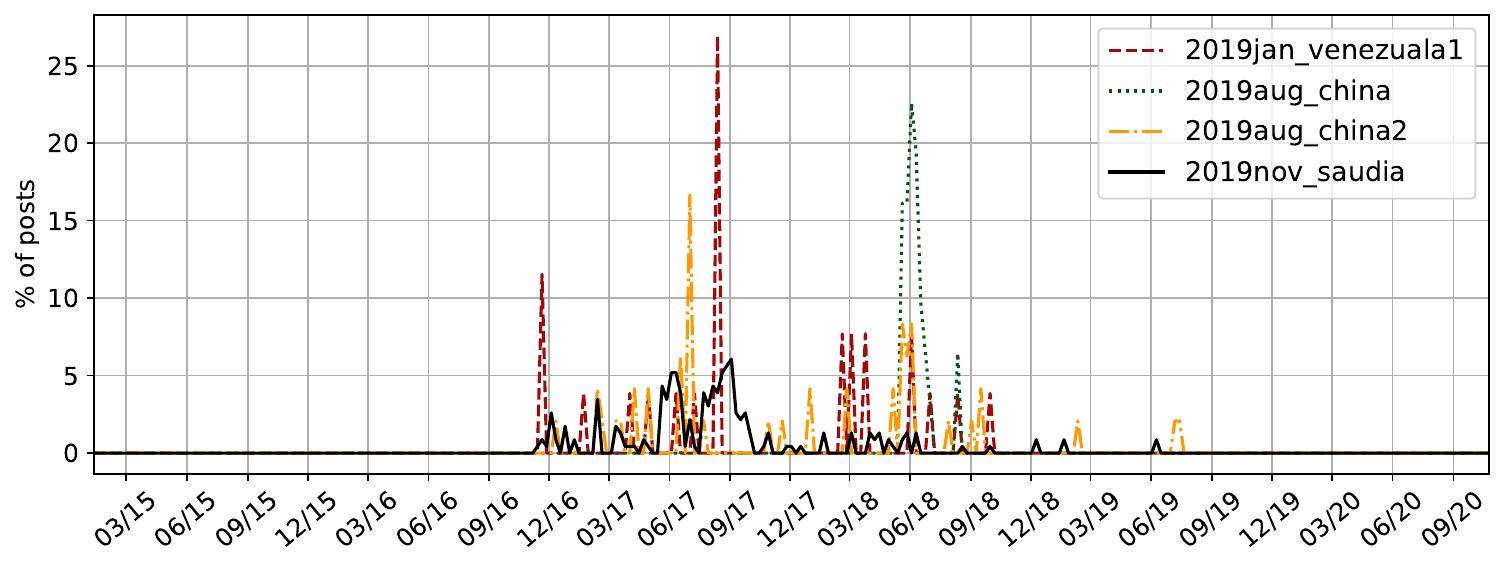}
		\caption{``Moments Internal Auth''}
		\label{fig:moments}
	\end{subfigure}
	
	\caption{The time graphs show the same applications being used around the same time period by different campaigns.}
	\label{fig:apps_coord}
\end{figure*}

\descr{Number of Sources.}
We randomly select 500 accounts from the Twitter 1\% data discussed in Section~\ref{sec:data} and collect their tweets from 2016 to 2020 using the Twitter Search API.\footnote{The Twitter API is no longer publicly available and does not show the tweet source.}
Figure~\ref{fig:sources_cdf} shows that, on average, a state-backed account uses more sources to post its messages than a regular account.
The maximum sources used by a regular user are 16 as compared to 115 for a state-backed account.
The discrepancy in the number of sources can be attributed to several factors.
For a regular user, it is likely that they might switch between different devices or platforms depending on their circumstances or contexts.
For example, they might use a mobile application while on the go and switch to a web client when using a computer.
However, in the case of a state actor, utilizing diverse sources can help manage their campaigns.
For example, using scheduled applications for sending messages that make the accounts appear real (e.g., ``advice messages''), and using other applications to spread the actual harmful content. 
It might also be advantageous to spread out application use when it comes to third-party applications, since abnormally high activity on fake sources might lead the sources to get flagged and suspended.
For example, the campaign \textit{2019jan\_venezuela1} used the fake application ``Twitter for\textvisiblespace\textvisiblespace Android'' primarily for retweets, where 4,061 out of the total 4,783 messages posted from the application were either retweets or replies to other tweets.

\descr{Simulating Legitimacy.}\label{sec:content}
As seen before, we find a trend of retweeting messages both inter- and intra-campaign.
On a similar note, we find that a number of messages ``unrelated'' to any specific agenda(s) are posted.
For example, below are ``advice'' messages that are verbatim shared by various unrelated campaigns (i.e., 2019jan\_iran, 2019aug\_china1, 2019jan\_russia, 2018oct\_ira, 2019nov\_saudia and 2019aug\_china2) suggesting access to a common knowledge base:

\begin{mdframed}[style=exampledefault, backgroundcolor=vlightgray]
	When a friend does something wrong, don't forget all the things they did right.
	\\ ---------
	
	\noindent When two people are meant for each other, no time is too long, no distance is too far, and no one can ever tear them apart.
	\\ ---------
	
	\noindent My past is my past, it made me who I am, I have no regrets, wouldn't change a thing. I just don't live there anymore.
	\\ ---------
	
	\noindent Just because I don't talk to you, or text you first, doesn't mean I don't miss you. I'm just waiting for you to miss me.
	\\ ---------
	
	\noindent I get jealous because I'm afraid someone is going to make you happier than I do.

\end{mdframed}

These quotes, presented as inspirational or uplifting content, might serve as camouflage for the dissemination of misinformation while building trust and credibility among their audience.
A similar behavior of posting cute dog pictures to appear legitimate and blend into the community has been shown in past research~\cite{saeed2022troll}.
The presence of positive and motivational content may divert attention from the deceptive nature of the underlying narratives and can also aid in avoiding detection~\cite{starbird2019disinformation}.
By appealing to emotions, these accounts might seek to amplify the impact of their disinformation, as individuals are more likely to share content that elicits emotional responses~\cite{vos2018spread}.

\descr{Timing and Coordination.}
We analyze various third-party applications and the times when they were used by various campaigns.
Figure~\ref{fig:apps_coord} highlights two applications being used by different campaigns.
The fake application ``Twitter for\textvisiblespace\textvisiblespace Android'' was actively used between 2013 and 2015 by four different campaigns.
Similarly, ``Moments Internal Auth'' shows various coordinated spikes between 2016 and 2019 for four different campaigns.
``Moments'' is an application that allows users to curate and present collections of tweets around specific topics or events.
It provides a way to aggregate tweets, including text, images, and videos, into a single, immersive narrative.
The plot in Figure~\ref{fig:apps_coord} shows an example of possible strategic coordination among these actors, potentially driven by shared objectives or alliances.
It is important to note that the accounts from each campaign were actively posting from other applications before and after these posts, further implying a strategic reason for using these applications during the specific timeframes (i.e., 2013-2014 in Figure~\ref{fig:android} and 2016-2019 in Figure~\ref{fig:moments}).
The synchronized usage hints at the existence of information-sharing networks or coordinated efforts to exchange tactics and strategies among state-sponsored actors.
It is also plausible that these custom applications were being marketed and sold to state actors, offering functionalities that align perfectly with their specific campaigns.

\descr{Pushing Agendas and Stylometry.}
We observe instances where different campaigns simultaneously push identical narratives.
The similarity can be attributed to geopolitical strategies, economic objectives, or ideological alignments.
When two or more campaigns have aligned goals, they may coordinate their disinformation efforts to amplify their desired message and enhance its impact.
It is also likely that in the realm of disinformation, successful strategies and tactics are emulated.
Below are some tweets in Arabic that are shared amongst three different Iranian campaigns (2019jun\_iran, 2019jun\_iran1 and 2019jun\_iran2), along with their translation (\textbf{NB:} Farsi, not Arabic, is the language spoken in Iran):

\begin{mdframed}[style=exampledefault, backgroundcolor=vlightgray]
	\textbf{Message 1:}
	\RL{من لم يكفّر الشيعة فهو كافر!}
	
	\noindent \textbf{Translation:}
	Whoever does not declare the Shiites to be infidels is an infidel!
	
	\noindent \textbf{Message 2:}
	\RL{ای شيعيان! شما به ما منسوب هستيد، پس مايه‌ی زينت ما باشيد نه مايه‌ی آبروريزی ما}

	\noindent \textbf{Translation:}	
	O Shiites! You belong to us, so be our adornment and not my disgrace

	\noindent \textbf{Message 3:}
	\RL{جز راه\_حسین ،باقی راه ها،بیراهه است}

	\noindent \textbf{Translation:}	
	Except the way of Hossein, the rest of the ways are misguided

	\noindent \textbf{Message 4:}
	\RL{نکاح حیوانات:رابطه باحیوانات…}
	
	\noindent \textbf{Translation:}	
	Marriage of animals: relationship with animals...
	
	\noindent \textbf{Message 5:}
	\RL{نکاح الحيوانات:معاشرةالحيوانات لتجنب الزنا}
	
	\noindent \textbf{Translation:}	
	\#Nekaah\_animals: Living with animals to avoid fornication!

\end{mdframed}

Sunni and Shia are two different sects of Islam that hold different beliefs and have ongoing tension; notably with respect to Iran/Saudi Arabia relations.
The first tweet calls out Shia Muslims as infidels, whereas tweets two and three are in favor of Shias.
In the context of the third tweet, Hossein is the grandson of the Muslim Prophet and is particularly respected in the Shia community.
The tweet calls all castes other than Shias misguided.
This is textbook controversy, which is the cornerstone of troll account behavior, i.e., accounts take both sides of the argument to cause strife~\cite{saeed2022troll}.
On the other hand, the last two tweets are examples of exaggerated claims and blatant disinformation while appealing to religious sentiment.
The author is promoting marriage (or ``Nekaah'' in Arabic) with animals to prevent the sin of fornication.

\descr{Takeaways.}
In a nutshell, we identify certain characteristics of state-sponsored campaigns, shedding light on their potentially coordinated nature and deceptive strategies.
We find seven key characteristics of state-backed operations: 1)~the extensive use of scheduling applications, 2)~impersonating third-party applications, 3)~leveraging a variety of sources, 4)~retweeting similar messages, 5)~simulating legitimacy, 6)~coordinating their timings, and 7)~pushing certain agendas.

Our findings suggest that there are shared behavioral patterns exhibited by state-sponsored accounts from different campaigns despite originating in different countries.
Recent work showed possible coordination amongst state actors in individual state-sponsored campaigns (e.g., retweeting behavior~\cite{wang2023state}).
We find a similar trend across different campaigns on a varying scale.
We analyze state-backed operations from various vantage points (e.g., information sharing behaviors and stylometry), in order to build a generalized feature fingerprint on how troll accounts disseminate information, communicate, and share content.
We find that state-backed operations exclusively use impersonated third-party applications that are not available on official app stores.
Multiple state-backed actors use these applications around the same timeframes.
The state-backed accounts also make use of scheduling applications and on average, use more sources than regular users to post content.
Although scheduling messages is traditionally associated with marketing campaigns, we observe that state-sponsored influence campaigns use this strategy too.
We also find stylometric similarities in the posted content: retweeting of similar messages and amplifying similar content at large scale.
The use of scheduling applications could serve as an easier method to scale such activities.
On a similar note, recent work points towards a posting behavior called \textit{copypasta}~\cite{vishnu2024track} networks where the same ``disinformation'' is duplicated in troll operations.
The stylometric similarities we observe in our dataset and the retweeting and duplication of content across state actors points towards similar behavior at scale.

Putting it all together, in the following sections, we translate our empirical findings into account-level features to demonstrate that state-backed accounts show clear differences from benign users for these metrics.

\subsection{Account-Level Characteristics}
We now map the identified campaign characteristics into account-level signals.
We prepare a list of 12 signals that capture various characteristics we have identified.
The signals incorporate posting styles in the form of stylometric features (i.e., unique words, sentence length, and others), group activity in terms of retweets and mentions, and the use of ``regular'' well-known sources (i.e.,``Twitter Web Client,'' ``Twitter for Android,'' ``Twitter for iPhone,'' ``Twitter for iPad,'' and ``Twitter Web App'') by the users.
To determine the possibility of leveraging these features to build a generalized troll account detection system, we first perform statistical tests on the populations to observe key differences.
In Table~\ref{tbl:stat}, we show the mean value of each of the given features derived from the campaign-level characteristics in a set of 1000 real world Twitter users obtained from Twitter 1\% data described in Section~\ref{sec:data} and all troll accounts belonging to state-backed operations.
We perform the Kolmogorov–Smirnov~\cite{kstest2022} test to determine whether the differences in scores are statistically significant.
For each feature, we perform the test and report the scores in Table~\ref{tbl:stat}, along with P-values.
We set the value of $\alpha$ to \pval, therefore, we only reject the null hypothesis if the P-value is $<$\pval.
Overall, we can reject the null hypothesis for all metrics, which means that for all samples, the difference in values are statistically significant.
Next, we aim to use these features along with others to build a machine learning classifier for detecting accounts from state-backed operations.

\begin{table}[t]
	\begin{center}
		\setlength{\tabcolsep}{2pt}
		\small
			\begin{tabular}{lllll}
				\toprule
				\textbf{Characteristic} & \textbf{Trolls}  &  \textbf{Real} & \textbf{KS} & \textbf{P-Val} \\ 
				\midrule
				Fraction of Messages by Regular Sources & 0.67 & 0.72 & 0.10 & \textbf{~$<$\pval} \\ 
				Cumulative Mentions per Tweet & 9.06 & 0.63 & 0.53 &\textbf{~$<$\pval} \\
				Fraction of Messages that are Retweets & 0.16 &  0.31 & 0.17 & \textbf{~$<$\pval}\\
				Number of Sources Used on Average & 2.50 & 2.26  & 0.11 & \textbf{~$<$\pval}\\
				Average Tweet Word Count & 15.18 & 12.66  & 0.14  & \textbf{~$<$\pval} \\
				Average Tweet Unique Words & 13.45 & 11.21  & 0.13  & \textbf{~$<$\pval} \\ 
				Average Tweet Stopword Count & 1.42 & 1.90 & 0.10 & \textbf{~$<$\pval} \\ 
				Average Tweet Punctuation Count & 2.89  & 2.50  & 0.11 & \textbf{~$<$\pval}\\ 
				Average Word Length & 5.44 & 4.64  & 0.15 & \textbf{~$<$\pval} \\ 
				Average Sentence Length & 12.33 & 9.66 & 0.17  & \textbf{~$<$\pval} \\ 
				Average Sentence Complexity & 52.52 & 40.14  & 0.21 & \textbf{~$<$\pval}\\ 
				Function to Non-function Words Ratio & 0.065 & 0.084 & 0.11 & \textbf{~$<$\pval} \\	
				\bottomrule
			\end{tabular}%
	\end{center}
	\caption{Statistical comparison between real users and state-sponsored accounts.}
	\label{tbl:stat}
\end{table}

\begin{table}[t]
	\begin{center}
		\setlength{\tabcolsep}{3.5pt}
		\small
			\begin{tabular}{lll}
				\toprule
				\textbf{No.} & \textbf{Feature}  &  \textbf{Novelty} \\ 
				\midrule
				1 & tweet\_count & \cite{echeverr2018lobo, forn2018holistic} \\
				& & \cite{ng2022botbuster, varol2017online} \\
				2 & account\_age & \cite{chu2012detect, davis2016bot}\\
				3 & no. of followers & \cite{echeverr2018lobo, forn2018holistic} \\
				& & \cite{liufake2018, ng2022botbuster} \\ 
				& & \cite{alex2021detect, varol2017online} \\ 
				4 & no. of following & \cite{forn2018holistic, ng2022botbuster}\\
				5 & language &  \cite{davis2016bot} \\
				6 & description\_length & \cite{liufake2018, alex2021detect} \\
				& & \cite{varol2017online} \\ 
				7 & description\_language & ours \\ 
				8 & cumulative\_mentions\_per\_tweet & \cite{varol2017online} \\ 
				9 & average\_tweet\_length & \cite{echeverr2018lobo} \\ 
				10 & retweet\_fraction & \cite{varol2017online} \\
				11-34 & percentage of tweets in a given hour & \cite{alzahbi200mach} \\
				35 & average\_tweet\_word\_count & \cite{varol2017online} \\
				36 & average\_tweet\_unique\_words & ours \\ 
				37 & average\_tweet\_stopword\_count & ours \\ 
				38 & average\_tweet\_punctuation\_count & ours \\ 
				39 & average\_word\_length & \cite{forn2018holistic} \\ 
				40 & average\_sentence\_length & \cite{forn2018holistic} \\ 
				41 & average\_sentence\_complexity & ours \\ 
				42 & function\_to\_non-function\_words\_ratio & ours \\
				43 & no\_of\_sources & \cite{echeverr2018lobo} \\ 
				44 & fraction\_of\_messages\_by\_fake\_sources & ours \\ 
				45 & fraction\_of\_messages\_by\_regular\_sources & ours \\			
				\bottomrule
			\end{tabular}%
	\end{center}
	\caption{System features.}
	\label{tbl:features}
\end{table}

\begin{figure*}[!t]
	\centering
	\includegraphics[width=0.925\textwidth]{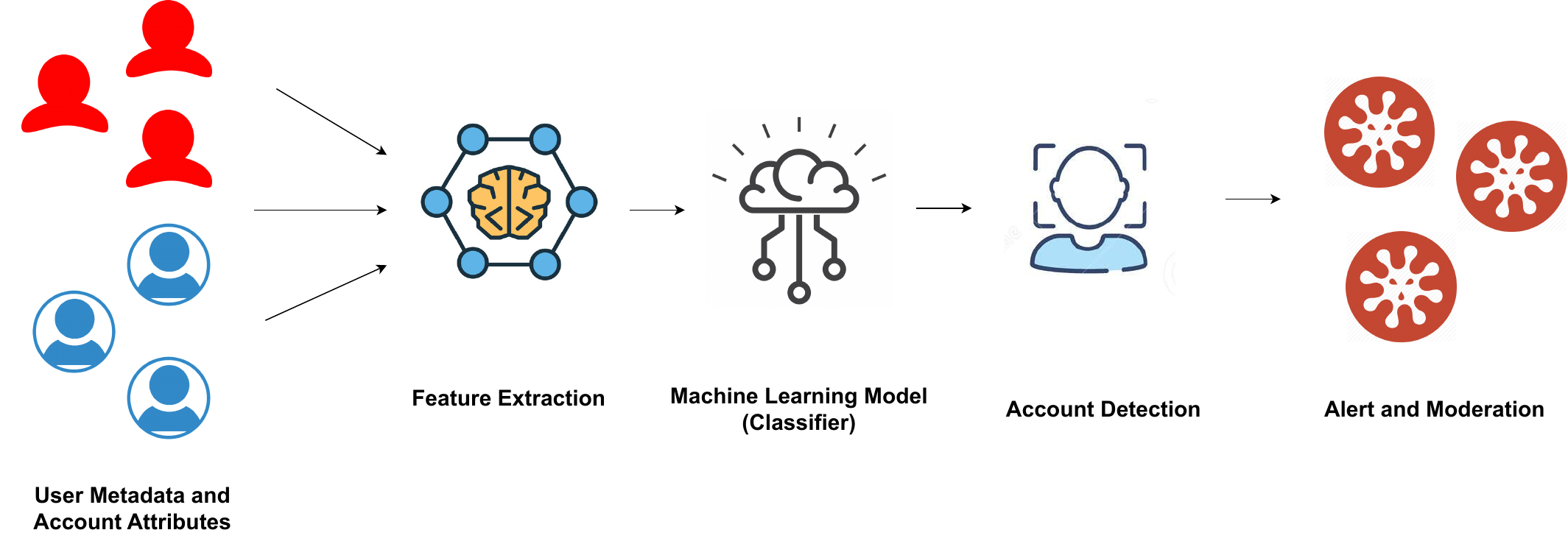}
	\caption{Overview of the system: two input streams are fed to the system: (1) a dataset of known state-sponsored malicious accounts and (2) a set of benign accounts. The system extracts the features of both sets of accounts. Next, a detection model is built using the identified features and used to detect unseen accounts in the wild. Finally, the system alerts users to potential malicious accounts in the wild, and those accounts can be moderated accordingly.}
	\label{fig:systempipeline}
\end{figure*}

\section{Building the Detection Model}
Based on our observations from Section~\ref{sec:charact}, we identify a set of features to train a supervised model to distinguish state-sponsored campaign accounts from legitimate Twitter accounts.
Figure~\ref{fig:systempipeline} lays out the entire pipeline of our system from feature extraction to finding potential state-backed accounts in the wild and alerting relevant authorities.
Unlike previous works~\cite{luceri2020detecting,saeed2022troll}, our model is designed to detect accounts from previously unseen campaigns, i.e., it can be trained on one campaign and used to detect accounts from other unknown campaigns.

\subsection{Feature Extraction}
To train a machine learning system, it is first important to identify and select relevant attributes or characteristics from the raw input data.
This process helps to transform the data into a suitable representation that the machine learning model can understand and learn from effectively.
In Table~\ref{tbl:stat}, we demonstrate the difference between state-backed actors and benign users for several features.
Since there are clear distinctions, we are positive that we can use certain features to discern between malicious and non-malicious users and build a machine-learning based classifier.
Therefore, in this section, we use those features and motivate several more that can be leveraged to build a detection system.
The complete list of features is shown in Table~\ref{tbl:features}.
Later on in Table~\ref{tbl:comp}, we show how removing certain types of features from the model (e.g., stylometric features) impacts the overall classifier performance and the importance for using all the features for the most optimal performance.

Overall, we divide the features into four main categories:

\descr{User Attributes.}
User metadata attribute features are valuable pieces of information that provide insights into the characteristics and behavior of users.

\begin{itemize}
	\item \textit{Tweet Count:}
	The number of tweets posted by an account provides insight into its activity level.
	State-sponsored accounts have an average tweet count of 2,483 as opposed to 5,433 of real users, with a KS-Score of 0.42 and P-Value$<$\pval, making the difference statistically significant.
	
	\item \textit{Account Age:}
	The age of an account can be indicative of suspicious behavior. According to past research, there is evidence of state-sponsored accounts being created around same the time frame or in batches~\cite{saeed2022troll}.
	
	\item \textit{Followers and Following:}
	State-sponsored accounts might follow each other to increase social proof and have a different follower-following ratio than a regular user.
	A state-sponsored account has 2,090 followers and 961 following while a regular user has 1,162 followers and 552, with a KS-Score of 0.17 and 0.16 respectively, with each P-Value$<$\pval, making the differences statistically significant.
	
	\item \textit{Language and Description Language:}
	On average, 37\% of state-sponsored accounts are non-English speaking, therefore we use the language of the account and its description as a feature.
	To model the language as a feature, we assign an integer value to each language.
	For example, ``English'' represented as ``en'' in the raw data would correspond to ``1.'' 
	In total, we consider 31 languages for our analysis, which are derived from both real-world users and state-backed accounts.
	
	\item \textit{Description Length:}
	The length of an account's description can provide insight into its authenticity.
	State-sponsored accounts have an average description length of 38.7 as opposed to 43.8 of genuine users, with a KS-Score of 0.078 and P-Value$<$\pval, making the difference statistically significant.
	
	\item \textit{Cumulative Mentions per tweet:}
	Twitter allows users to mention or tag other users in a tweet. 
	To mention someone in a tweet, the user includes their username preceded by the ``@'' symbol (e.g., ``@username [insert tweet text]'') within the tweet's text. 
	By calculating the cumulative average number of mentions per tweet, we identify patterns of interactions.
	Higher-than-average mentions as shown in Table~\ref{tbl:stat} indicates a coordinated effort to promote a specific agenda or target specific individuals or groups.
	For this feature, we count the instances of other users being mentioned by the target account.
	
	\item \textit{Average Tweet Length:}
	State-sponsored accounts may exhibit distinct patterns in tweet length. They have an average tweet length of 86.4 characters as compared to 68.7 characters of real world users, with a KS-Score of 0.194 and P-Value$<$\pval, making the difference statistically significant.
	
	\item \textit{Retweet fraction:}
	This feature indicates the fraction of tweets that are retweets.
	By analyzing the number of retweets made by the users, we identify the sharing rate.

\end{itemize}

\descr{Temporal Characteristics.}
These features are represented as a size 24 vector, where each entry depicts the percentage of messages in that hour from the 24-hour window.
Past research has shown evidence that troll accounts post during specific times of day, sometimes even as an office job~\cite{evgeny2017prof}.

\descr{Stylometry.}
Stylometry features allow us to capture linguistic nuances, such as vocabulary choices, sentence structure, punctuation, and grammatical patterns, that are characteristic of these accounts.
These features are derived from the differences we observe in Table~\ref{tbl:stat}.

\begin{itemize}

\item \textit{Average Tweet Word Count:}
By analyzing the average word count, the classifier can identify accounts that consistently produce unusually short or long tweets compared to real users.
Deviations from the expected average word count can be indicative of templated content commonly found in such accounts.

\item \textit{Average Tweet Unique Words:}
By calculating the average number of unique words in tweets, the classifier can identify the diversity in the vocabulary used by the accounts.

\item \textit{Average Tweet Stopwords Count:}
Stopwords are common words like ``and,'' ``the,'' or ``is'' that carry little semantic meaning.
Incorporating the stopwords count adds linguistic nuance to the classifier.

\item \textit{Average Tweet Punctuation Count:}
Punctuation usage can reveal certain writing styles or patterns associated with state-sponsored accounts.
Excessive use of punctuation marks, such as exclamation marks or ellipses, may indicate attempts to convey emotions or manipulate reader perception.

\item \textit{Average Word Length:}
State-sponsored accounts may utilize specific writing techniques, such as elongating words, to mask their writing style or bypass content filters.

\item \textit{Average Sentence Length:}
Sentence length can provide insights into the writing style and coherence of state-sponsored accounts.
By calculating the average sentence length, the classifier can identify accounts that show signs of unnatural content generation.

\item \textit{Average Sentence Complexity:}
By measuring average sentence complexity, such as the presence of complex sentence structures or syntactic constructions, the classifier can identify accounts that consistently produce content with a linguistic complexity indicative of malicious activity.
We use the Flesch Reading Ease~\cite{flesch} score to compute complexity.
It takes into account both the sentence length and the average number of syllables per word, providing a comprehensive measure of text complexity.
The score ranges from 0 to 100, with higher scores indicating easier-to-read text.

\item \textit{Function to Non-Function Words Ratio:}
Function words (e.g., articles, prepositions, and pronouns) and non-function words (e.g., nouns, verbs, and adjectives) have different roles in language.
State-sponsored accounts may exhibit patterns of overusing or underusing function words to manipulate or convey specific messages.

\end{itemize}

\descr{Source Features.}
We consider source features to be important because they provide valuable information about the origin and credibility of the content shared by state-backed accounts.
The following source features are included in the classifier:

\begin{itemize}

\item \textit{Number of Sources:}
State-sponsored accounts often use multiple sources to share content.
By analyzing the number of sources used in tweets, the classifier can identify accounts that exhibit an unusually high number of sources compared to real users.

\item \textit{Fraction of Messages by Fake Sources:}
State-sponsored accounts may purposefully share content from sources known for spreading misinformation or propaganda (e.g., fake third-party versions of original applications).
By tracking the number of sources with errors, the classifier can identify accounts that consistently share content from unreliable or deceptive sources.
We consider a source fake if it contains extra or leading spaces and all lower-case letters.

\item \textit{Fraction of Messages by Regular Sources:}
State-sponsored accounts use various scheduling applications and other third-party sources to post their messages.
Since the list of scheduling applications and new third-party applications is ever-expanding, we account for the fraction of messages made by accounts from ``regular'' well-known sources (i.e.,``Twitter Web Client,'' ``Twitter for Android,'' ``Twitter for iPhone,'' ``Twitter for iPad,'' and ``Twitter Web App'').

\end{itemize}

\subsection{Training the System}\label{sec:train}
We create a balanced dataset for training, with 500 accounts randomly selected from the Twitter 1\% dataset from Section~\ref{sec:data} as the negative class.
We use the Twitter Search API to retrieve their tweets.
To select accounts for the positive class, we use two campaigns with a high number of accounts, i.e., 2018oct\_ira with 3,608 accounts and 2018oct\_iran with 770 accounts.
We randomly select 500 accounts from both campaigns and use those as positive classes to train the system.
By using a balanced dataset, we ensure that the model is exposed to enough examples from both classes, reducing the risk of bias.
In real-world scenarios, the class of interest (the positive class) is often the minority class. 
By balancing the dataset, the model also becomes more sensitive to the minority class, allowing it to learn patterns and features specific to that class~\cite{jap2002class, yan2011class}.
We report the average result of both iterations in Table~\ref{tbl:class}.
For each iteration, we perform 10-fold cross-validation to train the system.
We experiment with four classifiers: K-Nearest Neighbors (KNN)~\cite{wein2009knn}, Decision Tree~\cite{saf1991decision}, Support Vector Machines (SVM)~\cite{suy1999least}, and Random Forest~\cite{liaw2002r}.
We evaluate the performance of each classifier based on accuracy, precision, recall, and F1-score.
We find that Random Forest classifier works the best, achieving an accuracy of 98.5\% and an F1-Score of 97.8\%.
Therefore, we select Random Forest for performing evaluation of our system and identifying accounts from unseen campaigns in the next section.

\begin{figure}[!t]
	\centering
		\includegraphics[width=0.5\textwidth]{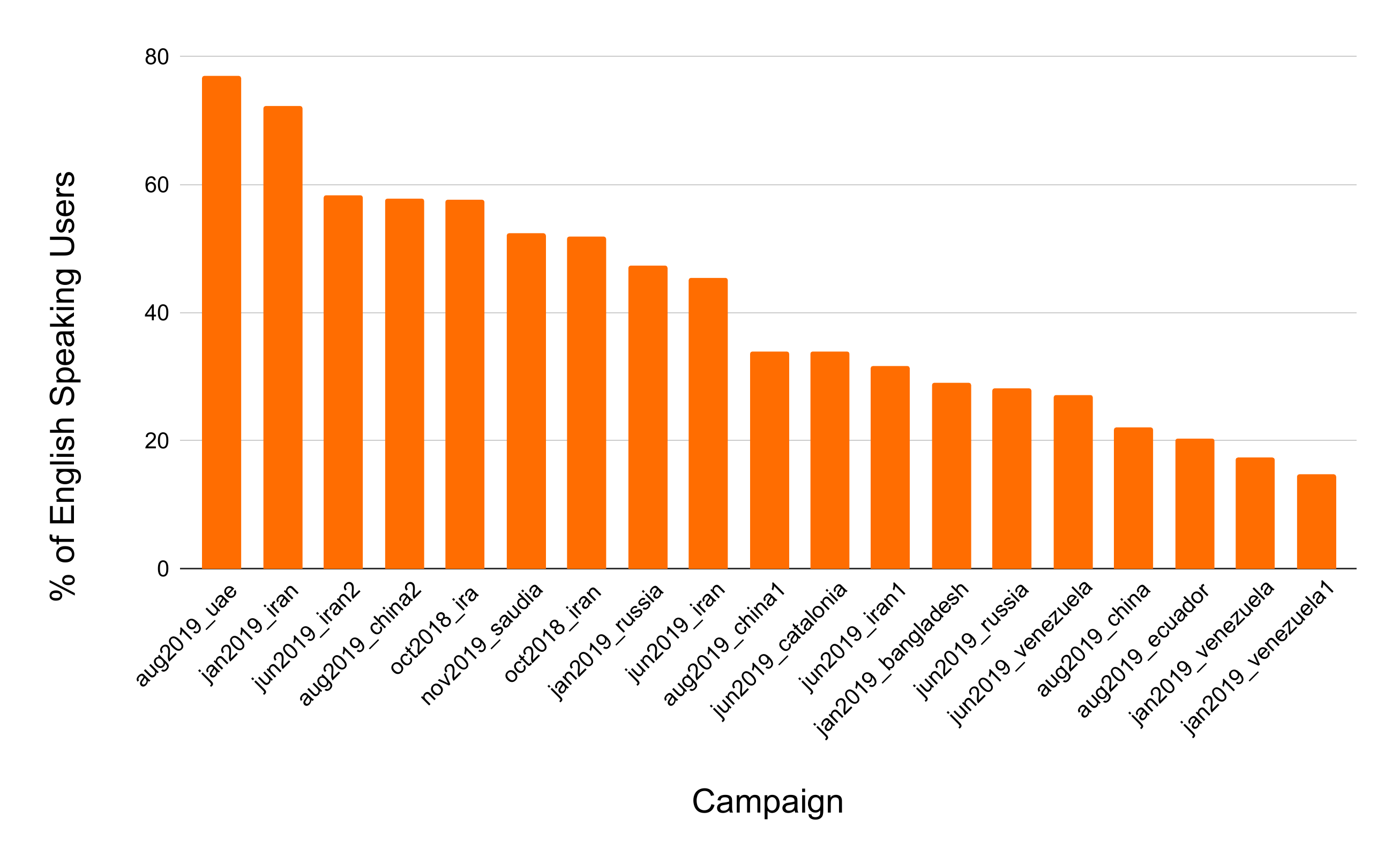}
	\caption{The graph shows the percentage of English-speaking users per campaign.}
	\label{fig:lang_camps}
\end{figure}

\begin{table}[t!]
	\begin{center}
		\setlength{\tabcolsep}{3pt}
		\small{
			\begin{tabular}{lrrrr} 
				\toprule
				{\bf Classifier} & {\bf Accuracy} & {\bf Recall} & {\bf Precision} & {\bf F1-Score} \\ 
				\midrule
				KNN	& 92.4\%	&	92.6\%	&	92.8\% & 92.7\% \\ 
				Decision Tree & 95.4\%    &  94.2\%     &  94.6\%     & 94.4\% \\ 
				Linear SVM	&	97.7\%	&	97.6\%	&	97.6\%  & 97.6\% \\ 
				\textbf{Random Forest}	& \textbf{98.5\%}	&	\textbf{97.6\%}	&	\textbf{97.9\%}  & \textbf{97.8\%} \\
				\bottomrule
			\end{tabular}
		}
	\end{center}
	\caption{Classification performance of our system using 10-fold cross-validation.}
	\label{tbl:class}
\end{table}

\begin{table}[t!]
	\begin{center}
		\setlength{\tabcolsep}{3pt}
		\small{
			\begin{tabular}{lrrrr} 
				\toprule
				{\bf Classifier} & {\bf Accuracy} & {\bf Recall} & {\bf Precision} & {\bf F1-Score} \\ 
				\midrule
				Metadata	& 95.4\%	&	96.2\%	&	96.3\% & 96.3\% \\ 
				Temporal & 91.4\%  &  92.9\%     &  84.8\%     & 88.9\% \\ 
				Stylometry	&	90.4\%	&	90.0\%	&	84.7\%  & 87.4\% \\ 
				Source	&	76.7\%	&	71.1\%	&	57.2\%  & 64.2\% \\ 
				\textbf{All}	& \textbf{98.5\%}	&	\textbf{97.6\%}	&	\textbf{97.9\%}  & \textbf{97.8\%} \\
				\bottomrule
			\end{tabular}
		}
	\end{center}
	\caption{Classification performance of each classifier component individually.}
	\label{tbl:comp}
\end{table}

\begin{figure*}
	\centering
	
	\begin{subfigure}[b]{0.49\textwidth}
		\centering
		\includegraphics[width=\textwidth]{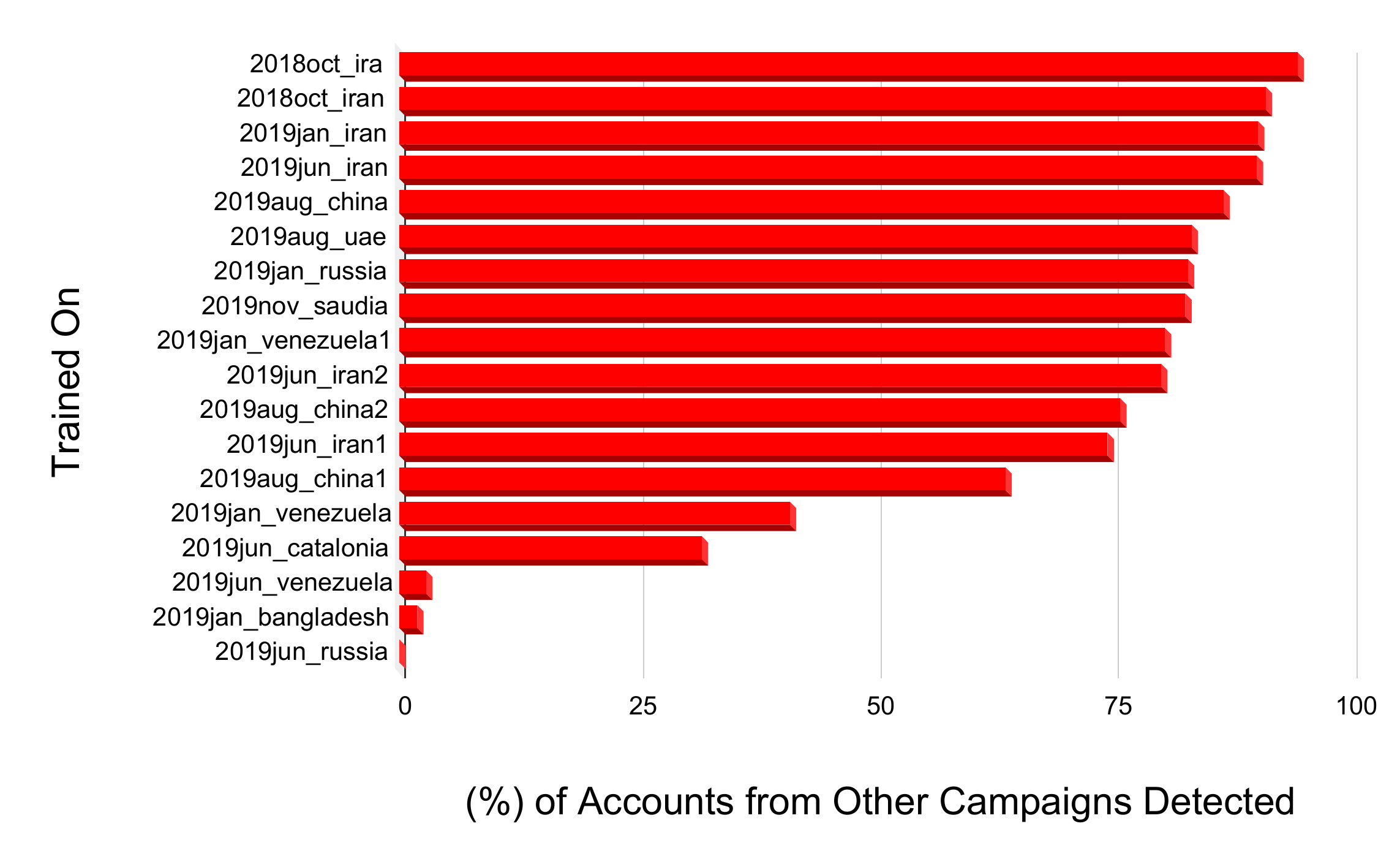}
		\caption{Training on Individual Campaigns and Evaluating on Others}
		\label{fig:eval_camps}
	\end{subfigure}
	\hfill
	\begin{subfigure}[b]{0.49\textwidth}
		\centering
		\includegraphics[width=\textwidth]{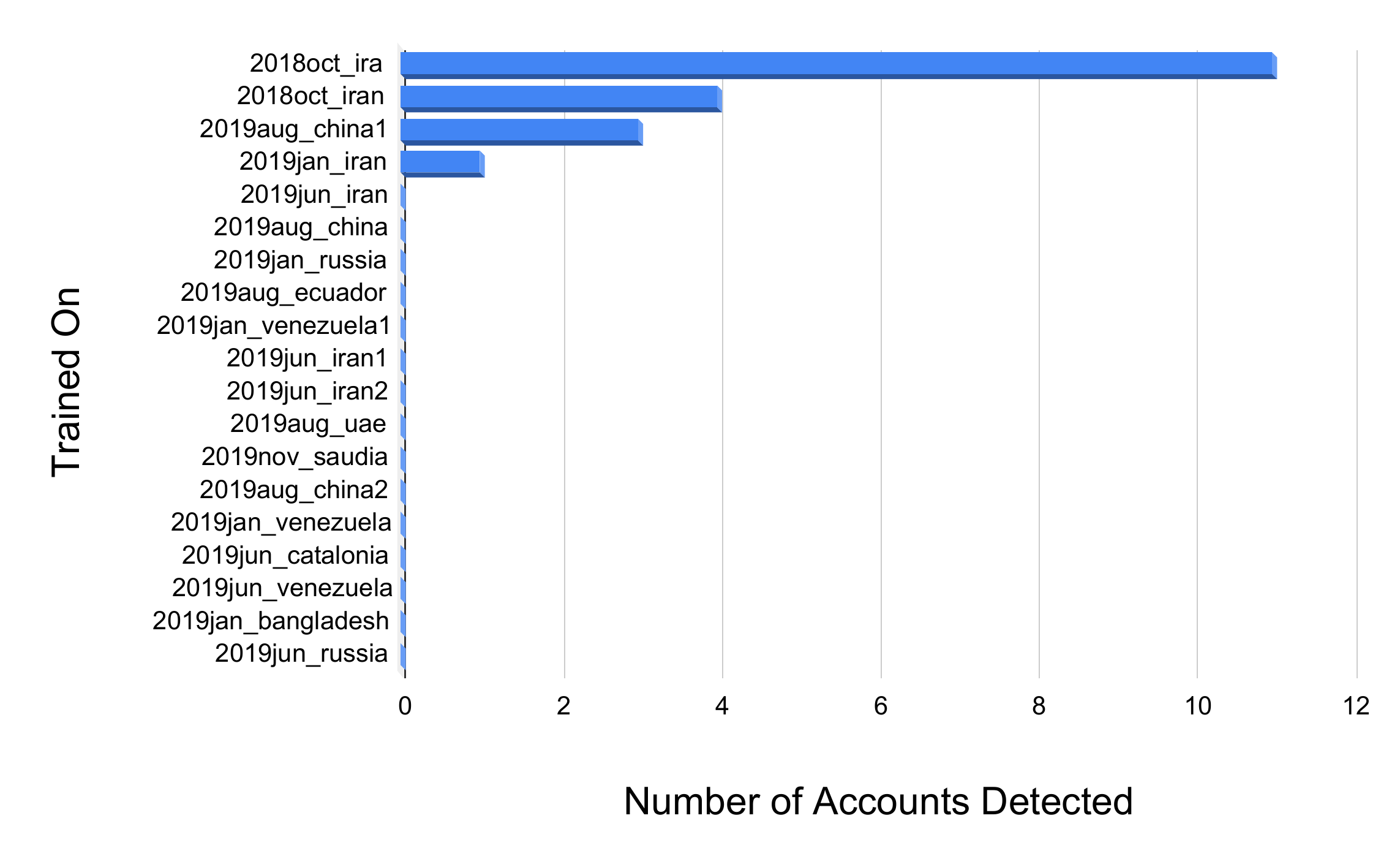}
		\caption{Training on Individual Campaigns and Evaluating on Real Users}
		\label{fig:eval_real}
	\end{subfigure}
	\caption{System performance and results.}
	\label{fig:eval}
\end{figure*}

\descr{Component Testing.}
We also test each component of our system independently and present the results in Table~\ref{tbl:comp}.
We find that, when considering individually, the metadata classifier performs the best, however it important to note that the stylometry and temporal classifier individually perform well.
By manually reviewing the misclassified accounts, we find that the source classifier is able to pick up accounts which are misclassified by the other three classifiers, adding overall nuance and signal towards prediction capabilities.
However, we achieve the best F1-score and accuracy when all components are put together.

\descr{Performance Using Unbalanced Dataset.} 
It is important to understand how a real-world imbalance could affect the performance of our system.
For this purpose, we use the 2019jun\_venezuela with 33 accounts as positive class for the system and 500 real-world users as negative class.
Due to limited data access, this is our best attempt to mimic a scenario with more examples of real-users.
We find that in this setting the accuracy of our model slightly increases, with an accuracy of 99.5\%, a precision score of 99.6\%, a recall of 99.6\% and an F1 score of 99.6\%.

\descr{Structural Imbalances in Dataset.}
To identify differences in state-sponsored actors and benign users, it is also important to account for structural imbalances in the dataset, for instance the language used by Twitter accounts.
Figure~\ref{fig:lang_camps} shows the percentage of English-speaking users per campaign, as determined by the ``language'' attribute in user metadata.
We see diversity among campaigns ranging from a small percentage of English-speaking users to nearly all users being English-speaking.
We find that the set of real users that we collect for our work consists of 26.7\% English-speaking users.
While language imbalance between state-sponsored accounts and benign accounts can introduce unintended bias in the model, we attempt to reduce the biases by evaluating our system by cross-validating on campaigns with different proportions of English-speaking users.

\descr{Feature Importance.}
We also examine the importance assigned to each feature by our machine learning model.
We compute the Gini importance~\cite{sci2023feat}, which assigns a value between 0 and 1 for each feature.
The more significant the feature for prediction, the higher its value.
We find that ``Retweet Fraction'' and ``Cumulative Mentions per Tweet'' are the best predictors, confirming our findings from Section~\ref{sec:charact} that state-sponsored accounts have distinct attributes in sharing, promoting and pushing their content.

Our system leverages several modalities to detect troll campaigns.
We propose a combination of user-level, temporal, stylometric, and source features for detection of state-backed accounts, where every component and its individual performance is highlighted in Table~\ref{tbl:comp}.
A portion of these features have been used in the past for different account detection systems as shown in Table~\ref{tbl:features}.
However, the combination of features we propose is obtained by studying campaigns from various state-actors emerging from different countries and comparing their behavior with benign users.
Our feature combination is unique and tailored towards influence operations.
We also propose novel features (e.g., impersonated application use and stylometry).
Past works have touched on some aspects of troll behavior, like loose coordination patterns~\cite{saeed2022troll}, retweeting and promoting of each other’s content~\cite{wang2023state} and duplicating or copy pasting text~\cite{vishnu2024track}.
However, our system provides a holistic approach towards detecting accounts, where the performance is best by using a combination of factors that make a successful state-backed campaign.

\section{Evaluation}
In the previous section, we show that the model performs well on previously seen campaigns.
Now, we use our system to detect accounts from unseen campaigns, i.e., we train the system individually on each campaign and detect the accounts from all the other campaigns.
We use the negative class accounts from Section~\ref{sec:train}, and accounts from each campaign are considered positive class.
Figure~\ref{fig:eval_camps} highlights the performance of our system.
When trained on the October 2018 IRA campaign, the system is able to detect roughly 94\% of the accounts from other campaigns, i.e., 24,247 out of 25,667 total accounts.
There are certain campaigns that train the system poorly, such as 2019jun\_russia.
However, it is important to note that these campaigns have very few accounts for the classifier to train on, like 2019jun\_russia consists of only three accounts.
With enough examples of the positive class, the classifier is able to detect more accounts from unseen campaigns.

\descr{Estimating False Positives.} We also test our system on a separate set of real-world users randomly selected from the Twitter 1\% dataset.
From Figure~\ref{fig:eval_real}, the highest number of accounts flagged is eleven out of 629 when the system is trained on 2018oct\_ira campaign.
Therefore, we estimate the upper-bound false detection rate at roughly 1.75\%, which is the worst-case scenario. 
For the majority of campaigns our system shows a 0\% false detection rate.

\begin{table}[t]
	\begin{center}
		\setlength{\tabcolsep}{3.5pt}
		\small
			\begin{tabular}{ll}
				\toprule
				\textbf{Flagged Accounts}  &  \textbf{Other Accounts} \\ 
				\midrule
				admtvosenlucha &si \\
				followers &quiero \\
				\RL{اللهم} &vida \\
				\RL{الله} & hoy \\
				stats & día \\
				igualdad & gracias \\
				todas & mejor \\
				partes & siempre \\
				8m & así \\
				unfollowers & amor \\
				twitter & love \\
				come & tan \\
				join & bien \\
				media& solo \\
				policy & dios	\\		
				\bottomrule
			\end{tabular}%
	\end{center}
	\caption{Top 15 words computed using TF-IDF for accounts that were flagged and not flagged by our system in the wild.}
	\label{tbl:tfidf}
\end{table}

\subsection{Detection in the Wild}\label{sec:wild}
We search the Twitter 1\% data from 2017 to find messages made from sources that are popularly used by the coordinated campaigns identified in Section~\ref{sec:charact}.
We find 2,696 accounts that posted one or more messages from the application ``Twitter for\textvisiblespace\textvisiblespace Android''.
We then run our system to identify accounts that might belong to state-backed operations.
Our system marks 116 out of all the accounts.

\descr{Comparison of Flagged vs Other Accounts.}
Now, we compare the content features of accounts flagged by our system with the ones that are not flagged to look for linguistic signals.
We use TF-IDF (Term Frequency - Inverse Document Frequency) for this task.
A word's importance to a document in a collection or corpus is meant to be reflected by the TF-IDF statistic, which is a numerical measurement.
Since some words are used more frequently than others overall, the TF-IDF value rises according to the number of times a word appears in the text and is offset by the number of documents in the corpus that contain the term.
We use it to ensure the selection of words that are both important within a specific document and distinctive in the overall dataset.
We compute the top 15 words using TF-IDF for tweets from both sets of accounts: 1) flagged and 2) non-flagged or other accounts.
The results are shown in Table~\ref{tbl:tfidf}.
Most words from the accounts not flagged by the system are generic in nature.
The presence of religious terms like ``\<اللهم>'' and ``\<الله>'' (which mean ``God'' in Arabic) may suggest the exploitation of religious sentiments or potential attempts to influence religious communities through propaganda or divisive content.
On the other hand, we observe terms like ``si,'' ``quiero,'' ``vida,'' ``hoy,'' and ``día'' in the non-flagged dataset, which are common Spanish words related to daily life. 
These terms and others like ``gracias,'' ``mejor,'' and ``siempre'' show typical social interactions rather than promoting specific agendas.

\section{Case Studies}\label{sec:case}
In this section, we show various case studies from the accounts identified by our system in Section~\ref{sec:wild} which highlight their involvement in state-backed operations.
We do so to present additional evidence that these accounts are potentially malevolent and operate in ways that recognized disinformation campaign accounts do.

\descr{Retweeting Similar Agendas.}
As discussed in Section~\ref{sec:retweet}, a common strategy accounts in state-backed operations use is to retweet a large chunk of messages.
A major reason to retweet certain messages can be to boost a given narrative and increase the visibility and exposure of selective content to a larger audience.
Retweeting each other's messages also helps create an echo chamber effect by giving the illusion of widespread support or consensus.
To the outside eye, repeated retweets from seemingly independent accounts give the impression that multiple sources support and validate the information being shared, thereby enhancing the perceived credibility of the messages.
It is in the best interest of malicious actors to create the perception that their propaganda has gained traction and popularity.
By retweeting each other's content, these accounts can also inflate engagement metrics, such as retweet counts and likes, while drowning out or overshadowing genuine voices and opinions.
When targeting an important event, such as a political event, it is specifically advantageous for a malicious actor to flood the platform with repetitive content, which can aid in spinning a desired narrative and creating the illusion of overwhelming support for their agendas.
This can discourage or intimidate genuine users from expressing dissenting views, ultimately stifling open and balanced discussions.
The following examples contain three retweets that were made by accounts identified by our system as well as those from known disinformation campaigns (i.e., 2019jan\_iran and 2019jan\_venezuela1).

\begin{mdframed}[style=exampledefault, backgroundcolor=vlightgray]
	\textbf{Message 1:}
	RT [REDACTED]: Al Assad: El ataque químico fue "ciento por ciento fabricado" y los reportajes son "falsos" 
	
	\noindent \textbf{Translation:}
	RT [REDACTED]: Al Assad: The chemical attack was "one hundred percent fabricated" and the reports are "false"
	
	\noindent \textbf{Message 2:}
	RT [REDACTED]: La Patria de Bolívar rechaza la enfermiza obsesión de Luis Almagro contra Venezuela, exigimos respeto absoluto. Somos libres…
	
	\noindent \textbf{Translation:}	
	RT [REDACTED]: The Homeland of Bolívar rejects the sick obsession of Luis Almagro against Venezuela, we demand absolute respect. We're free…

	\noindent \textbf{Message 3:}
	RT [REDACTED]: Este 19 de abril decimos con Alí: yo no me quedo en casa pues al combate me voy!. Vamos a la Calle, vamos a la Batalla, vamo…

	\noindent \textbf{Translation:}	
	RT [REDACTED]: This April 19 we say with Ali: I'm not staying at home because I'm going to fight! Let's go to the Street, let's go to the Battle, let's go…
\end{mdframed}

The first message is regarding a ``chemical attack'' about which the mainstream media reports are flawed according to the tweet.
Casting suspicion on authority is one of the major themes revolving around disinformation campaigns, as has been seen in past research (e.g., about mistrusting the government on COVID-19 vaccines)~\cite{mel2022stan}.
The second message is a pro-Venezuela tweet, which is also a common theme amongst campaigns: to pick a side of the story and spread the message as much as possible.
The last retweet is a call-to-action and can be considered an instigation to start protests.
It is one of the key components of disinformation campaigns and is aimed at causing disruptive behavior.
Past works have named this as \textit{Rapid disinformation attacks}, whereby, disinformation is unleashed quickly to sabotage a crucial event (e.g., right before an important election)~\cite{john2020ai}.

\descr{Advice Messages.}
Another important aspect of accounts involved in disinformation campaigns is their ability to slip under the radar and appear ``legitimate.''
To do so, these accounts often employ deceptive tactics to gain the trust of their target audience.
One such tactic is posting ``advice'' messages, which are designed to provide seemingly helpful or insightful information while subtly promoting a specific agenda.
By posting advice messages, state-sponsored accounts might aim to establish themselves as trustworthy and knowledgeable sources.
These messages also target the user's emotions and aspirations.
In the past, various techniques have been used to fulfill this objective, ranging from posting cute dog pictures to sharing jokes~\cite{saeed2022troll}.
As discussed in Section~\ref{sec:content}, the commonly observed strategy is to post ``advice'' messages that are unrelated to any specific agenda(s) being pushed by the campaign.
Following are two examples of such messages that were posted by accounts identified by our system and those in a known campaign (i.e., 2019jan\_venezuela1):

\begin{mdframed}[style=exampledefault, backgroundcolor=vlightgray]
	\textbf{Message 1:}
	Una persona que realmente te conoce es alguien que ve el dolor en tus ojos, mientras los demás creen que sonríes.
	
	\noindent \textbf{Translation:}
	A person who really knows you is someone who sees the pain in your eyes, while others think you smile.
	
	\noindent \textbf{Message 2:}
	La diferencia entre "puedo" y "no puedo" es solo de una palabra y una cuestión de actitud; si hay ganas, todo se puede.

	\noindent \textbf{Translation:}	
	The difference between "I can" and "I can't" is only one word and a matter of attitude; if there is desire, everything is possible.
	
\end{mdframed}

It is interesting to observe that these messages are copied and pasted verbatim, highlighting the possibility of a corpus of such messages that are sent through these accounts just to make them look like normal users.
Both the messages are in Spanish, and therefore, we provide an English translation for them.
The first message is more uplifting, and the next one is motivational.
By posting advice messages, they can blend seamlessly with the larger user base, making it challenging to distinguish them from authentic accounts.
This tactic might also lead to accounts remaining active for extended periods and continue their propaganda efforts undetected.

\section{Related Work}
In this section, we examine previous studies that have investigated fake social media accounts and disinformation campaigns.

\descr{Detection of Fake Accounts.}
A wide variety of research looks at detecting fake accounts on social media.
Some efforts have been made to determine features found commonly in fake accounts, e.g., a disproportionate friend-to-follower ratio or the similarity of posted content~\cite{benevenuto2010detecting,stringhini2010detecting}.
More sophisticated systems, such as the one proposed by Yang et al., discovered more resilient features in fake accounts that are difficult for adversaries to evade (e.g., average neighbor's followers and following rate)~\cite{yang2011free}.
Chu et al. propose a classifier to distinguish between a human, a cyborg, and a bot using a system built on various components (e.g., entropy and spam detection)~\cite{chu2012detect}.
Ghosh et al. delved into the phenomenon of \textit{link farming}, which spam accounts use to amass a large number of followers~\cite{ghosh2012understanding}, while Wang et al. analyzed user click patterns to create user profiles and employed supervised and unsupervised learning techniques to detect fake accounts~\cite{wang2013you}.
Viswanath et al. utilized Principal Components Analysis (PCA) to identify patterns among extracted features from spam accounts~\cite{viswanath2014towards} while Egele et al. concentrated on detecting compromised legitimate accounts, finding that regular users exhibit consistent habits over time, and any sudden deviations from these patterns indicate a compromise~\cite{egele2015towards}.
Davis et al. propose \textit{BotorNot}, a system that uses over 1000 features (including Temporal, User, and Friend features) to detect whether an account is real or a bot~\cite{davis2016bot}.
Another distinction found between regular and fake accounts is \textit{social connections}.
In this direction, Danezis et al. applied a Bayesian Inference approach to detect compromised accounts by identifying communities with similar characteristics~\cite{danezis2009sybilinfer}, while Cai et al. partitioned social networks into communities and sought out ones that displayed inconsistent connections with the rest of the network~\cite{cai2012latent}.

A separate line of research focuses on the synchronized behavior of fake accounts, which are commonly controlled by a single entity.
Cao et al. propose \textit{SynchroTrap}, a detection system that groups malicious accounts based on their synchronized actions and timing~\cite{cao2014uncovering}, while Stringhini et al. introduce \textit{EvilCohort}, a system that identifies sets of social network accounts utilized by botnets by examining communities of accounts accessed by shared IP addresses~\cite{stringhini2015evilcohort}.

Closely related to our work is a system proposed by Alhazbi et al.~\cite{alzahbi200mach}.
The authors use a set of behavioral features to detect state-sponsored troll accounts on Twitter.
They train and evaluate on a set of Saudi trolls disclosed by Twitter in 2019, with an overall classification accuracy of up to 94.4\%.
However, in our work we identify features that are common across several campaigns spanning multiple countries.
Our work also uses features from different modalities (e.g., temporal and stylometric) as we demonstrate the interplay of different features to capture different modus operandi used by troll accounts.
We also demonstrate the efficacy of our system by training on many different campaigns and identifying unseen state-sponsored accounts, thus generalizing the utility of our system beyond a single campaign.
Another system proposed by Fornacciari et al.~\cite{forn2018holistic} uses six groups of features based respectively on the analysis of writing style, sentiment, behaviors, social interactions, linked media, and publication time to detect state-sponsored troll accounts on Twitter.
The system, TrollPacifier, uses a neural network model and achieves a classification accuracy of 95.5\%.

\descr{Disinformation Campaigns.} 
Numerous studies have examined the role of social bots in the spread of political disinformation~\cite{bessi2016social,ferrara2017disinformation, ferrara2016rise,varol2017online}.
This body of work demonstrates that bots are capable of large-scale public opinion manipulation, which may have an impact on important political events, such as election results.
Mihaylov et al. show that there are two types of accounts involved in spreading disinformation: independent actors and those who are financially incentivized to propagate specific messages~\cite{mihaylov2016hunting}.
Steward et al. study Russian-sponsored troll accounts participating in the Black Lives Matter (BLM) discussion on Twitter and show that they infiltrated both left- and right-leaning communities with the objective of promoting particular narratives~\cite{steward2018examining}, while Ratkiewicz et al. use machine learning techniques to identify the spread of false political information on Twitter~\cite{ratkiewicz2011detecting}.
Zannettou et al. conduct a variety of studies examining state-sponsored troll accounts operating on Twitter and Reddit between 2014 and 2018 and evaluate their effectiveness in disseminating content across various platforms and web communities ~\cite{zannettou2019characterizing, zannettou2019disinformation,zannettou2019let}.
They show that the accounts involved in such campaigns are often created in \textit{waves} and later present a pipeline that focuses on studying the images shared by these accounts on Twitter.
Similarly, Hegelich et al. analyze the utilization of 1.7K Twitter bots during the Russia-Ukraine conflict and uncover a range of behaviors exhibited by these bots, such as attempts to conceal their identity, promotion of specific topics through hashtags, and the retweeting of messages with particularly engaging content~\cite{hegelich2016are}.
More recently, Wang et el.~\cite{wang2023state} find some inter-state coordination patterns in state-backed operations on Twitter highlighting that influence campaigns attract greater attention than baseline information operations, however their contribution is a measurement instead of a detection system.

\section{Discussion and Conclusion}
In this paper, we study a publicly available dataset of state-sponsored coordinated campaigns on Twitter.
These campaigns are geared towards spreading disinformation, trolling, and other disruptive and malicious behaviors.
We show that there are patterns in the way these campaigns operate (e.g., use of scheduling services for automation and impersonating third-party applications).
From our findings, we derive a machine-learning based model that can identify accounts from unseen campaigns.
Unlike past research, we take a step towards building systems that can perform cross-campaign detection.
We show that it is possible to build systems that are tailored towards the intricate nature of state-sponsored campaigns.
Additionally, we find various instances of potential inter- and intra-state coordination (e.g., coordinated posting patterns, shared applications and copy-pasted messages), hinting at a larger market or shared modus operandi used by different state actors.
I.e., it is highly unlikely that these campaigns are completely homegrown.
In this direction, some past research has pointed towards black markets for reusable political disinformation bots~\cite{ferrara2017disinformation}.
We believe that it is important for future research to look at disinformation and influence campaigns through a larger lens.

\descr{Resilience to Evasion.}
Our multi-factor analysis approach is a key component of our detection system's resilience to evasion.
The system considers multiple indicators of disinformation spread, including user metadata attributes, linguistic patterns, and temporal behaviors making the system less susceptible to evasion targeting specific features.
Disinformation campaigns attempting to evade detection by using one specific technique may still exhibit other suspicious patterns that can be captured by the system. 
This multi-factor analysis helps reduce the effectiveness of evasion strategies and enhances the system's overall robustness.

For example, an attacker might attempt to avoid detection by minimizing one of the popular tactics (e.g., the use of scheduling applications to automate messages).
Although it might avoid detection, it would make their campaigns much less effective since the messages would have to be manually sent and require much more effort, reducing the efficiency (and likely the efficacy) of the campaign to a large degree.

Another evasion strategy could be to reduce the frequency of retweeting or sharing each other's messages.
However, this will also reduce the efficacy of their operations since the reach and visibility of the disinformation content will be diminished.

\descr{Implications for the Design of Disinformation Detection Systems.}
There are several implications for the design of new safety features for social media, especially platforms like Twitter, Threads, and Mastodon that can be derived from our findings.
First, our analysis provides a crucial step toward identifying the characteristics of inauthentic accounts that spread disinformation to enable automated detection (machine learning based) of troll accounts in the future.
In particular, we establish the distinct ways in which these accounts operate, strategize and often present themselves to appear human-like.
Understanding how accounts belonging to state-sponsored disinformation campaigns attempt to mimic legitimate users can help inform behavioral models.
These models can track account behavior over time and compare it to established patterns of disinformation spreaders.
By continuously monitoring and analyzing user behavior, detection systems can adapt to evolving tactics, identifying previously unseen strategies. 

Next, we use a combination of features that can help future systems distinguish troll accounts from legitimate ones.
Rather than relying on individual signals alone, detection systems can combine multiple indicators to assess the likelihood of an account engaging in disinformation campaigns.
This holistic approach increases the robustness and accuracy of the detection process.

It is also worth noting that while machine learning models play a crucial role in automated detection, human expertise is invaluable in fine-tuning and validating the system's output.
Combining the insights gained from our research with human analysts' expertise can act as a force multiplier for the detection system.

Another implication for social media platforms is to be more cognizant of the idea of third-party agencies aiding different state actors in mass-producing disinformation.
The presence of copy-pasted content, coordinated posting, third-party clients shared by multiple actors and other nefaroius methods being used should inform the development of future systems countering disinformation.

\descr{Limitations and Future Work.}\label{sec:lim}
Like any other malicious actor, disinformation campaigns are likely to adapt to counter-detection efforts.
Since these campaigns are ever-evolving and are now becoming \emph{smarter} and more \emph{outsourced}, thereby expanding their attack vector by using a variety of sources (e.g., blogs, websites, fake articles and dedicated accounts)~\cite{bilva2023dis}, it is possible that the detection system's efficacy might drop against novel strategies in a long enough time period. 

We advance the field of combating online disinformation by identifying emerging trends and tactics used by disinformation campaigns and using those to build a detection system that can effectively perform campaign-agnostic detection.
However, we also argue that in order to maintain a robust defense, continuous monitoring and updates are essential to effectively address the changing climate of disinformation with the presence of LLM tools and the possibility of widespread disinformation~\cite{zellers2019defending}, as is true for many information security related problems (ranging from fuzzers to spam detection to browser fingerprinting).

Another constraint in our study is the selection process of real accounts used for training our classifier in Section~\ref{sec:train}.
We cannot guarantee with absolute certainty that the chosen set of accounts does not contain any accounts from a disinformation campaign.

In Section~\ref{sec:wild}, we test the system on a set of 2,696 accounts and identify 116 accounts that potentially belong to state-backed operations.
However, we do not have definite proof of these accounts being malicious, but we illustrate their resemblance with the activity of accounts belonging to known campaigns through case studies in Section~\ref{sec:case}.
A related limitation is the discontinuation of the Twitter API, leaving us unable to test our system on a larger, empirical dataset.
Unfortunately, this is a problem that will be faced by all social media research moving forward, but at the same time, our dataset is ``clean'' in that it predates the widespread availability of generative text models like ChatGPT.

Future research could delve deeper into the mechanisms of coordination, investigate the objectives and motivations behind such convergence, and explore the impact of shared application usage on the effectiveness of disinformation campaigns.
It is also imperative for researchers to look at disinformation campaigns from a broader perspective and understand the commonalities between them to build better detection systems.
Additionally, understanding the role of platform policies, algorithmic biases, and countermeasures in mitigating the influence of state-sponsored disinformation remains an important area for further investigation.

\descr{Acknowledgments.} This work was supported by the National Science Foundation under Grants CNS-1942610, CNS-2114407, CNS-2114411, CNS-2247867, and CNS-2247868. 
Any opinions, findings, and conclusions or recommendations expressed in this Report are those of the PI and do not necessarily reflect the views of the NSF.

\bibliographystyle{abbrv}
\bibliography{refs}

\end{document}